\documentclass[a4paper,11pt]{article}
\usepackage{latexsym}
\usepackage{feynmf}		
\usepackage{graphicx}
\def\ba{\begin{eqnarray}}
\def\ea{\end{eqnarray}}

\def\lb{\label}
\def\be{\begin{equation}}
\def\ee{\end{equation}}

\begin{document}
\baselineskip0.25in
\title{About the isocurvature tension between axion and high scale inflationary models}
\author{M. Estevez $^{1}$\thanks{e-mail:septembris.forest@hotmail.com}\  and O. Santill\'an $^{2}$\thanks{e-mail: firenzecita@hotmail.com, osantil@dm.uba.ar}\, \,.\date {}
\\
\\ \\
{\small $^1$ CONICET, IFIBA}\\
{\small Ciudad Universitaria Pab. I, Buenos Aires 1428, Argentina.}\\
\\
{\small $^2$ CONICET--Instituto de Investigaciones Matem\'aticas Luis Santal\'o IMAS,} \\
{\small Ciudad Universitaria Pab. I, Buenos Aires 1428, Argentina.} }
\maketitle

\begin{abstract}
The present work suggests that the isocurvature tension between axion  and high energy inflationary scenarios may be avoided by considering a double field inflationary model involving  the hidden Peccei-Quinn Higgs and the Standard Model one. Some terms in the lagrangian we propose explicitly violate the Peccei-Quinn symmetry but, at the present era, their effect is completely negligible. The resulting mechanism
allows a large value for the axion constant, of the order $f_a\sim M_p$,  thus the axion  isocurvature 
fluctuations are suppressed even when the scale of inflation $H_{inf}$ is very high, of the order of $H_{inf}\sim M_{gut}$. This numerical value is typical in Higgs inflationary models.  An  analysis about topological defect formation in this scenario is also performed, and it is suggested
that, under certain assumptions, their effect is not catastrophic from the cosmological point of view.

 \end{abstract}
\section{Introduction}
The axion mechanisms are an attractive solution to the CP problem in QCD \cite{axion1}-\cite{axionn}. In their simplest form, the axion $a$ is identified as a Nambu-Goldstone pseudo scalar corresponding
to the breaking of the so called Peccei-Quinn symmetry. This  is a $U(1)$ global symmetry which generalizes the standard chiral one. There exist models in the literature for which this symmetry breaking
takes place in a visible sector \cite{pq}, or in a hidden one \cite{ksvz}-\cite{dfsz}. In particular, the KSVZ axion scenario \cite{ksvz} postulates the existence of a hidden massive quark $Q$, which
behaves as a singlet under the electroweak  interaction. This quark acquires its mass throughout a Higgs mechanism involving a neutral Peccei-Quinn field $\Phi$. Since this quark does not interact with the photon and with the massive $Z$ and $W$ bosons, the corresponding Nambu-Goldstone pseudo boson $a$ is not gauged away. Standard current algebra methods show that the mass of this axion $a$ is inversely proportional to the scale of symmetry breaking $f_a$ \cite{bardeen}.
There are phenomenological observations which fix this scale $f_a>10^{9}$GeV \cite{Solar}. This lower bound is required for suppressing  the power radiated in axions by the helium core of a red giant star to the experimental accuracy level.

Besides these constraints, there are estimates that suggest the upper bound $f_a<10^{12}$GeV \cite{abbott}. This bound insures that the present axion density is not higher than the critical one. 
The idea behind this bound is the following. The standard QCD picture is that the axion potential is flat until the universe temperature is close to $T_{qcd}$. Below this temperature
there appears an induced periodic potential $V(a)$, and the axion becomes light but massive. A customary assumption is that the axion is at the top of the potential $V(a)$ at the time where this transition 
occurs. When the Hubble constant is of the same order as the axion mass this pseudo scalar falls to the potential minimum and starts coherent oscillations around it. The initial amplitude, which correspond to a maximum, is $A\sim f_a$ and thus, the energy stored at by these oscillations is of the order
$E\sim A^2 m_a^2$. The authors  of \cite{abbott} analyzed the evolution of these oscillations to the present universe and found that the axion energy density today would be larger than the critical one $\rho_c \sim 10^{-47}$GeV$^4$ unless the bound $f_a<10^{12}$GeV takes place.

The axion has many interesting properties from the particle physics point of view. However, there exist some cosmological problems about them, specially in the context of inflationary scenarios. These problems depend on whether the Peccei-Quinn symmetry is broken during, at the end, or after inflation \cite{lesgur}. If the symmetry breaking takes place after inflation, then  axionic strings are formed when the temperature falls down below the temperature $f_a/N$, with $N$ is the integer characterizing the color anomaly of the model. These strings produce relativistic actions, which only acquire masses when the universe temperature is comparable to $\Lambda_{qcd}$. At this point these axions 
become a considerable fraction of dark matter. Constraints on axion model related to this axion production by radiating strings and string loops have been studied in \cite{harari}-\cite{harari3}. There is the possibility that the breaking occurs at the end of the transition, for which the formation of the strings is qualitatively different \cite{Lyth}. 

An alternative to this  problem is that the symmetry is broken at the end of inflation. The topological defects that arise in this situation are qualitatively different than the strings discussed above and, to the best of our knowledge, they have not been studied yet \cite{lesgur}.  

A further possibility is that the breaking takes place before inflation, which implies that  the strings are diluted away due to the rapid expansion of the universe. 
This softens the axionic domain wall problem. Scenarios of this type takes place when $H_{inf}$ is below the value $2\pi f_a/N$. In this case the relic  density is suppressed by a factor $\exp(N_{e})$ with $N_e$ the number of e-folds that take place
between the symmetry breaking and the end of the inflation. For $N_e$ large enough, the suppression may be effective, and the density of such relics will be negligible today \cite{lesgur}.

The last possibility discussed above is attractive from the theoretical point of view. However, for this realization of symmetry breaking, the bound $10^{9}$GeV$<f_a<10^{12}$GeV is in tension with high energy inflationary models. This is due to the fact that the axion is effectively massless at the inflationary period and, for any massless scalar (or pseudo scalar) field  $a$ present during inflation, there will appear quantum fluctuations with nearly scale invariant  spectrum of the form
$$
<\delta a^2(k)>=\bigg(\frac{H_{inf}}{2\pi}\bigg)^2\frac{2\pi^2}{k^3}.
$$
This is an standard result, which implies that isocurvature perturbation corresponding to the field $a$ is given by \cite{Lyth}
$$
S_{CDM}=r_a\frac{\delta a}{a}=r_a\frac{H_{inf}}{2\pi f_a},
$$
with $r_a$ the fraction of $a$ particles in the present CDM. When this result is applied to axions, the observational constraints on $S_{CDM}$ \cite{planck} together with the axion window $10^{9}$GeV$<f_a<10^{12}$GeV  put constraints on $H_{inf}$ of the form $H_{inf}<10^{7}-10^{10}$GeV. For this reason, there is  an special interest in relaxing the axion window $10^{9}$GeV$<f_a<10^{12}$GeV, since otherwise the existence of a solution to the CP problem may  enter in conflict with the existence of high scale inflation, where high scale means $H_{inf}>10^{10}$GeV.

A well known example of these high scale models is the Higgs inflationary scenario \cite{higgsinf}. This model is very attractive, since introduces a single parameter to the Standard Model. This dimensionless parameter, denoted  by $\xi$, has a numerical value $\xi\sim 5.10^{4}$ and describes the non minimal coupling between the Higgs and the curvature $R$.  This minimality generated a vivid interest in the subject.The scale at the end of inflation for this scenario is of the order of $H_{inf}\sim 10^{15}$GeV, which is not far to the GUT scale. Thus, if it is assumed that the symmetry breaking takes place at inflation, one should find mechanisms for which initially $f_a\sim 10^{17}-10^{19}$GeV for avoiding the isocurvature problem.  This scale is essentially the Planck mass, and violates the bound in\cite{abbott} by seven magnitude orders. The present paper is related to this problem.

A valid approach for solving the isocurvature problem is to assume that $f_a$ is of the order of the Planck mass today. The bound $f_a<10^{12}$GeV assumes that at the beginning of the QCD era the axion is at the top of its potential. Thus an axion constant $f_a\sim M_p$ can be introduced in the picture if at the beginning of this era the axion already has rolled to a lower value by some unknown dynamics. If the  axion mass during the inflationary and the reheating periods is not zero, and in fact very large, the axion may roll to the minimum in an extremely short time before the QCD era. There exist some mechanisms in the literature where this aspect is discussed \cite{dvali}-\cite{dvali2}. Further interpretations of these problems and an update of the cosmological constraints may be found in \cite{wilczek} and references therein.

In the present work, a double Higgs inflationary mechanism \cite{double1}-\cite{doublen} involving the ordinary Higgs and the KSVZ Peccei-Quinn field will be considered.  It is argued here that the KSVZ field falls to the minima inside the inflationary period, in such a way that the topological defects are diluted away.
The present model contains some explicitly Peccei-Quinn symmetry violating terms which induces an small axion mass at the early universe. The key point is that when the terms induced after the QCD transition are added to the original potential coming from inflation, the result is the interchange between the the maxima and the minima. It is suggested that these initial terms are irrelevant at the present era, but they may induce the axion to sit in the point $a\sim 0$ during the universe evolution, thus avoiding the bound $f_a<10^{12}$GeV. In addition, several cosmological constraints on the parameter of the model are also discussed in detail. There exist a related work combining double Higgs inflation with the DFSZ axion \cite{doublen}, and a comparison between that work and the present one will be presented in the conclusions. 

The present work is organized as follows. In section 2 some known models dealing with the isocurvature problem are briefly discussed. This description is exhaustive, but the facts described there are the ones that inspire our work.  In section 3 a mechanism for avoiding the isocurvature problem is described in detail. This mechanism is a convenient modification of the double Higgs inflationary scenarios adapted to our purposes.  Section 4 contains a discussion  about the formation of the topological defects in our model. It is argued there that the contribution of topological defects is not relevant and
they axion emission do not overcome the critical density. Section 5 contains some variations of the model, and describe in detail the relevance of some of the parameters. Section 6 contains a discussion of the results 
and comparison with the existing literature.
 
\section{Preliminary discussion}
\subsection{General scenarios  related to the isocurvature problem}
Before to turn the attention into a concrete model, it may be instructive to describe some known mechanism which deal with the isocurvature problem. The following discussion 
is not complete but it is focused in some facts to be applied latter on. 

A not so recent approach to the isocurvature problem is to consider some non renormalizable interactions between the inflaton $\chi$ and the Peccei-Quinn field $\Phi$. For instance, in a supersymmetric context, there is no symmetry preventing a term of the form $
\delta K=\frac{1}{M_p^2}\chi^\dag \chi \Phi^\dag\Phi$  \cite{randall1}-\cite{randall2}, which can be present at the Planck scale. At inflationary stages, where the field $\chi$ is the dominating energy component, these terms induce an effective coupling of the form
$\Delta V(\Phi)=c H^2\Phi\Phi^\ast$ , with $c$ a dimensionless constant \cite{randall1}-\cite{randall2}. Furthermore, when supergravity interactions are turned on, a generic expression for these corrections may be of the form 
$\Delta V(\Phi)=H^2 M_p^2 f\bigg(\frac{\Phi}{M_p}\bigg)$ , with $f(x)$ a model dependent function \cite{randall1}-\cite{randall2}. Thus, for a high scale inflation, these corrections may be considerable since the value of $H$ is large. On the other hand, depending on the model, the sign of these corrections may be positive or negative. For instance, the authors \cite{Takahashi-Yamada} consider soft supersymmetry breaking terms which lead to an effective potential of the form
$$
V(\Phi)=m_\Phi^2 \Phi\Phi^\ast-c_H H^2\Phi \Phi^\ast-\bigg(a_H\lambda H \frac{(\Phi\Phi^\ast)^2}{4M_p}+c.c\bigg)+\lambda^2 \frac{(\Phi\Phi^\ast)^3}{4M^2_p},
$$ 
with $a_H$, $c_H$ and $\lambda$ the effective parameters of the model. Note that the sign of the second term is opposite to the first one. These models assume the presence of  physics beyond the Standard Model, but the addition of such terms can induce a large expectation value for $\Phi$ at the inflationary period which suppress isocurvature perturbations. Further details about this mechanism may be found in the original literature.

The scenarios discussed above fulfill  the bound $10^9$GeV$<f_a<10^{12}$GeV and postulate that the isocurvature fluctuations are suppressed due to a dynamical effective  symmetry breaking scale $f_a\sim M_p$
which evolves to a  lower value later on. A variant for these scenarios is to consider assume that $f_a\sim M_{p}$, and therefore the bound  $10^9$GeV$<f_a<10^{12}$GeV is in fact violated. This will be the approach to be employed for the authors in the following. Scenarios of this type may be realized  if there is some dynamical process previous to the QCD transition epoch that forces the axion $a$ to be much below than the top of the potential $a\sim f_a$. These possibilities were discussed for instance in \cite{dvali}-\cite{dvali2}, where the authors present several contribution to the axion mass $m_a$ in the early universe which are negligible today. These models require corrections that come from physics that comes from supersymmetric scenarios or even string theory ones.

Some scenarios that go in those directions are the ones in \cite{Masperi1}-\cite{Masperi4}. These models are considered in the context of electroweak strings with axions and their applications to bariogenesis, and introduce effective corrections to the axion mass of the form
\be\lb{Masperi2}
V(a,H)=\frac{\lambda}{4}(H H^\dag-v)^2+\bigg(m_\pi^2 f_\pi^2+f(H H^\dag-v)\bigg)\bigg[1-\cos\bigg(\frac{a}{f_a}\bigg)\bigg].
\ee
The function $f(x)$ is not known, but it is assumed that $f(0)=0$. This implies that, when the Higgs $H$ field is at the minimum, there are no correction to the axion mass, i.e,  $m_a\sim m_\pi f_\pi/f_a$ \cite{axion3}. Thus the low energy QCD picture is unchanged in the present era.

The corrections  (\ref{Masperi2}) suggest the following solution to the isocurvature problem. The corrections $f(H H^\dag-v)$ and the term $m_\pi^2 f_\pi^2$ may have opposite sign, in such a way that the sign of the term multiplying the function $\cos(a/f_a)$ is negative. In this case the point $a=0$ is now a  maximum instead a minimum. By assuming, as customary, that the axion is initially at the top of the potential, it is concluded the initial value may be $a\sim 0$. Furthermore,  when the inequality $H_a(t)>m_a(t)$ is satisfied during the universe evolution, the axion is frozen in an small neighbor $a\sim0$. If in addition, there is a time for which the value of $m^2_\pi f^2_\pi$ has  absolute value larger than $f(H H^\dag-v)$, then the sign of the potential changes, but the axion did not evolve and is still is near $a\sim 0$. This violates the hypothesis \cite{abbott} and thus the bound $f_a<10^{12}$GeV is eluded since the initial axion value at the QCD transition era is not $a\sim f_a$ but  instead $a\sim 0$.

\subsection{Generalities about double Higgs inflationary models}

The  discussion given above suggests that the corrections to the axion mass (\ref{Masperi2}) may be important for softening the tension between high energy inflationary and axion models. However, the authors
\cite{Masperi1}-\cite{Masperi4} did not give a complete explanation of the dynamical origin of such a mass term.  Nevertheless, it  is clear from  (\ref{Masperi2}) that, when the Higgs is at not at the minima, there are some violations of the  Peccei-Quinn symmetry. Otherwise, the axion would be massless. Thus, it is necessary to include  Peccei-Quinn violating terms in our scenario but simultaneously, it should be warranted that their effects are not important at present times.  A possibility is to employ some version of double Higgs inflationary models \cite{double1}-\cite{double3}, when some small but explicitly breaking Peccei-Quinn terms are allowed into the picture. These models however, do not consider a singlet Higgs, and this type of Higgs are essential in axion models. For these reason, it will be convenient  to describe the main features of double Higgs inflationary models, in order to adapt them to our purposes later on.

In general, the double Higgs scenarios contains two scalar fields doublets $\Phi_1$ and $\Phi_2$ with a non zero minimal coupling to the curvature $R$. This coupling is described by two parameters denoted by $\xi_1$, $\xi_2$ and $\xi_3$. The lagrangian for such model in the Jordan frame is given by \cite{double1}-\cite{double3}
$$
\frac{L_J}{\sqrt{-g_J}} = \frac{R}{2}+\Big(\xi_1|\Phi_1|^2+\xi_2|\Phi_2|^2+\xi_3\Phi^\dagger_1 \Phi_2+{\rm c.c.}\Big)R
-\left|D_\mu\Phi_1\right|^2-\left|D_\mu\Phi_2\right|^2-V_J\left(\Phi_1,\Phi_2\right).
$$
Here the covariant derivative $D_\mu$ corresponds to the electroweak interactions, but it may be allowed to correspond to another type of interactions if gauge invariance is respected.
The potential  $V_J(\Phi_1,\Phi_2)$ is the generic two Higgs one described in detail in \cite{gunion}-\cite{haver}, namely
$$
V(\Phi_1,\Phi_2) =  -m^2_1|\Phi_1|^2-m^2_2|\Phi_2|^2+\Big(m^2_3\Phi^\dagger_1\Phi_2+{\rm. c.c.}\Big)
+\frac{1}{2}\lambda_1|\Phi_1|^4+\frac{1}{2}\lambda_2|\Phi_2|^4+\lambda_3 |\Phi_1|^2|\Phi_2|^2
$$
\be\lb{guha}
+\lambda_4 \left(\Phi^\dagger_1\Phi_2\right)\left(\Phi^\dagger_2\Phi_1\right)+\left[ \frac{1}{2}\lambda_5\left(\Phi^\dagger_1\Phi_2\right)^2+\lambda_6\left(\Phi^\dagger_1\Phi_1\right) \left(\Phi^\dagger_1\Phi_2\right) + \lambda_7 \left(\Phi^\dagger_2\Phi_2\right) \left(\Phi^\dagger_1\Phi_2\right)+{\rm c.c.} \right].
\ee
In the following, the choice of dimensionless parameters will be such that always $\xi_3=0$ and $\lambda_6=\lambda_7=0$. The remaining non vanishing  parameters $m_i$ and $\lambda_i$ are assumed to be real. The lagrangian given above is expressed in units for which $M_p=1$, but  the dependence on this mass parameter will be inserted back later on.

The scalar doublets of the model may be parameterized as
\begin{equation}\lb{para}
\Phi_1=\frac{1}{\sqrt{2}}\left(\begin{array}{l} 0 \\ h_1 \end{array} \right) \, , \quad
\Phi_2=\frac{1}{\sqrt{2}}\left(\begin{array}{l} 0 \\ h_2 e^{i\theta} \end{array} \right).
\end{equation}
As for the standard Higgs inflationary model, the physics of the double Higgs model is clarified  by performing a Weyl transformation $g^J_{\mu\nu}=g^E_{\mu\nu}/\Omega^2$ with an scale factor
$\Omega^2\equiv 1+2\xi_1|\Phi_1|^2+2\xi_2|\Phi_2|^2$. By  assuming that the fields have large values  $\xi_1 h^2_1+\xi_2 h^2_2>> 1$ and by making the following field redefinitions
\be\lb{fr}\chi=  \sqrt{\frac{3}{2}}\log(1+\xi_1 h^2_1+\xi_2 h^2_2),\qquad r =  \frac{h_2}{h_1},\ee it is found that the previous action can be expressed in the following form \cite{double3}
$$
\frac{L_E}{\sqrt{-g_E}} \sim  \frac{R}{2} - \frac{1}{2}\left(1+\frac{1}{6}\frac{r^2+1}{\xi_2 r^2+\xi_1}\right)(\partial_\mu \chi)^2-\frac{1}{\sqrt{6}}\frac{(\xi_1-\xi_2)r}{\left(\xi_2 r^2+\xi_1\right)^2}(\partial_\mu \chi)(\partial^\mu r)
$$
\be\lb{homi}
-\frac{1}{2}\frac{\xi^2_2 r^2+\xi^2_1}{\left(\xi_2 r^2+\xi_1\right)^3}(\partial_\mu r)^2-\frac{1}{2}\frac{r^2}{\xi_2r^2+\xi_1}\left(1-e^{-2\chi/\sqrt{6}}\right)(\partial_\mu\theta)^2-V_E(\chi,r,\theta).
\ee
The potential energy (\ref{guha}) should be expressed in terms of the redefined fields as well. In the following, the quartic terms are assumed to be predominant and the quadratic ones, proportional to $m_i$ will be neglected. 
The resulting potential energy is approximated by
\begin{equation}
\label{poto}
V_E(\chi,r,\theta)= \frac{\lambda_1+\lambda_2 r^4+2\lambda_L r^2+2\lambda_5 r^2\cos(2\theta)}{8\left(\xi_2 r^2+\xi_1\right)^2}\,\left(1-e^{-2\chi/\sqrt{6}}\right)^2,
\end{equation}
with the definition $\lambda_L\equiv \lambda_3+\lambda_4$. The subscript $E$ will be omitted from now and on, and it will be understood that all the variables are related to the Einstein frame. 

It is convenient to remark  that the distinction between Jordan and Einstein frames is important at the early universe. However
for large times the scale factor $\Omega^2\sim 1$ and this distinction is not essential \cite{higgsinf}. 

Now, the potential for the quotient field $r$ defined in (\ref{fr}) is given by \cite{double1}-\cite{double3}
\begin{equation}\label{rpot}
V(r)\simeq \frac{\lambda_1+\lambda_2 r^4+2\lambda_L r^2}{8\left(\xi_1+\xi_2 r^2\right)^2}.
\end{equation}
The kinetic term for such field is not canonical, and scales as $\sqrt{\xi}$. The canonically normalized field is very massive \cite{double2} and is not slow rolling. Thus $r$ fastly stabilizes  at the minimum $r_0$ and the effective potential of the neutral Higgses and the pseudo-scalar Higgs becomes
\begin{equation}\label{efecto}
V(\chi,\theta) \simeq \frac{\lambda_{\rm eff}}{4\xi^2_{\rm eff}} \left(1-e^{-2\chi/\sqrt{6}}\right)^2 \left[1+\delta \cos(2\theta)\right],
\end{equation}
where $\delta \equiv\lambda_5r^2_0/\lambda_{\rm eff}$,
$\xi_{\rm eff}\equiv\xi_1+\xi_2 r^2_0$ and
$\lambda_{\rm eff}\equiv\left(\lambda_1+\lambda_2 r^4_0+2\lambda_L r^2_0\right)/2$,
with the finite value of $r^2_0$ given by
\begin{equation}\label{erre}
r^2_0=\frac{\lambda_1\xi_2-\lambda_L\xi_1}{\lambda_2\xi_1-\lambda_L\xi_2}.
\end{equation}
In this case, the effective non-minimal coupling and the effective quartic coupling are
$$
\lambda_{\rm eff} =  \frac{\lambda_1\lambda_2-\lambda^2_L}{2} \frac{\lambda_1\xi^2_2+\lambda_2\xi^2_1-2\lambda_L\xi_1\xi_2}{\left(\lambda_2\xi_1-\lambda_L\xi_2\right)^2},\qquad
\xi_{\rm eff} = \frac{\lambda_1\xi^2_2+\lambda_2\xi^2_1-2\lambda_L\xi_1\xi_2}{\lambda_2\xi_1-\lambda_L\xi_2}.
$$
In these terms, the inflationary vacuum energy becomes \cite{double1}-\cite{double3}
\be\lb{vaco}
V_0=\frac{\lambda_1\lambda_2-\lambda^2_L}{8\left(\lambda_1\xi^2_2+\lambda_2\xi^2_1-2\lambda_L\xi_1\xi_2\right)}.
\ee
Note that $U(\theta)$ becomes flat (or trivial) when $\delta=0$.

In the discussion given above, the quadratic terms of the potential (\ref{guha}) have been neglected. Howver, these terms are relevant in our model, since they are decisive in the evolution of the axion field. The quadratic potential in the Einstein frame with the variables (\ref{fr}) is given by
\be\lb{cuadr}
V_q=\frac{M_p^4}{2(\xi_1+\xi_2 r^2)^2h_1^2}(-m_1^2-m_2^2 r^2+2 m_3^2 r\cos\theta),
\ee
where the dependence on $M_p$ was inserted back.

\section{An scenario for avoiding the axion isocurvature problem}

In view of the formulas given above, it is tempting to define $\theta=a/f_a$ from where an axion $a$ emerges. Recall that the standard axion QCD potential goes as $V(a)\sim 1-\cos (a/f_a)$ while, if $m_3^2>0$ in (\ref{cuadr}), the term $\cos (a/f_a)$ in the potential (\ref{cuadr}) is positive. Thus the early and the QCD contributions are of opposite sign. This will be essential in our scenario, by the reasons discussed below the formula (\ref{Masperi2}). In addition, the potential (\ref{efecto})  also looks like an axion one, but with the opposite sign if $\delta$ is positive. This non zero value for the potential makes perfect sense, since the parameter
$\delta\sim \lambda_5$ and the coupling induced by a non zero $\lambda_5$ violates explicitly the Peccei-Quinn symmetry of the model. When the dependence on $M_p$ is inserted back into (\ref{efecto}), the induced potential becomes
\begin{equation}\label{efectos}
V(\chi, a) \simeq \frac{M_p^4\lambda_{\rm eff}}{4\xi^2_{\rm eff}} \left(1-e^{-2\chi/M_p\sqrt{6}}\right)^2 \left[1+\delta \cos\bigg(\frac{2a}{f_a}\bigg)\right].
\end{equation}
Thus the potential gets factorized as $V(\chi, a)=V(\chi)U(a)$ with $V(\chi)$ the standard Higgs potential in the transformed frame.  Furthermore the function $V(\chi)$
coincides with the potential  for the Higgs in the single inflation model \cite{higgsinf}. For larger times the conformal factor $\Omega^2\sim 1$, $H\sim \chi$ and a pion description of the strong interactions is possible. Then $V(a,\chi)$ becomes equal to the potential in the Jordan frame. The resulting expression clearly resembles (\ref{Masperi2}) as well.

Despite these resemblances with axion physics, the application of the formulas given in the previous section to the KSVZ scenario is not straightforward. First of all, the standard double Higgs extensions of the 
Standard Model contain two Higgs doublets $\Phi_1$ and $\Phi_2$ with hyper charge $Y=1/2$, otherwise the potential (\ref{guha}) would not be gauge invariant.
Instead, the KSVZ axion model contains the Standard Model Higgs $\Phi$ and hidden complex Peccei-Quinn scalar, which we will denote $\varphi$,  which is neutral under the electroweak interaction.
Thus direct application of the  previously presented results may enter in conflict with gauge invariance.

The drawbacks described above will be avoided as follows. First of all, a new real neutral scalar field $\beta$ will be introduced in the picture.
The lagrangian to be considered is now
$$
\frac{L_J}{\sqrt{-g_J}} = \frac{M_p^2}{2}R+\Big(\xi_1|\Phi|^2+\xi_2|\varphi|^2+{\rm c.c.}\Big)R
-\left|D_\mu\Phi\right|^2-\left|\partial_\mu\varphi\right|^2-\frac{1}{2}\left|\partial_\mu\beta\right|^2-V_J\left(\Phi, \varphi, \beta\right).
$$
Here the covariant derivative $D_\mu$ corresponds to the electroweak interactions, as before, and only the Higgs $\Phi$ participates in this interaction.
The potential  $V_J(\Phi, \varphi, \beta)$ is a modification of (\ref{guha}) and is given by
\be\lb{guha2}
V_J(\Phi, \varphi, \beta) =  
\frac{1}{2}\lambda_1(|\Phi|^2-v^2_1)^2+\frac{1}{2}\lambda_2(|\varphi|^2-f^2_a)^2
+\frac{1}{2}m_\beta^2 \beta^2 + \bigg(\frac{1}{2}\lambda_5|\Phi|^2 \varphi^2+ \mu \beta \varphi+{\rm c.c.}\bigg).
\ee
This potential is gauge invariant and it is assumed that $v_1\sim 246$GeV while $f_a$ is not far from the Planck scale. Both Higgs fields are parameterized as 
\begin{equation}\lb{para2}
\Phi=\frac{1}{\sqrt{2}}\left(\begin{array}{l} 0 \\ h \end{array} \right) \, , \quad
\varphi=\frac{1}{\sqrt{2}} \rho e^{i\theta} .
\end{equation}
In the following the case $\xi_2=0$ will be considered by simplicity. By defining the standard single Higgs inflation variable \cite{higgsinf}
\be\lb{aeb}
\chi =  \sqrt{\frac{3}{2}} M_p\log \bigg(1+\frac{\xi_{1} h^{2}}{M_p^2}\bigg),
\ee
the resulting lagrangian becomes
\be\lb{kined}
\frac{\mathcal{L}_{E}}{\sqrt{-g_{E}}}=\frac{M_p^2}{2}R-\frac{1}{2}(\partial_\mu \chi)^2-\frac{e^{-\sqrt{\frac{2}{3}}\frac{\chi}{M_p}}}{2}(\partial _{\mu} \rho)^{2}+\frac{e^{-\sqrt{\frac{2}{3}}\frac{\chi}{M_p}}}{2}\rho ^{2}(\partial _{\mu} \theta)^{2}+\frac{e^{-\sqrt{\frac{2}{3}}\frac{\chi}{M_p}}}{2}\left|\partial_\mu\beta\right|^2-V_{E}(h,\rho,\theta),
\ee
where now
\be\lb{i}
V_{E}(h,\rho,\theta)=e^{-2\sqrt{\frac{2}{3}}\frac{\chi}{M_p}}\bigg(
\frac{ \lambda_{1}}{8} (h^{2}-v_1^2)^2+\frac{ \lambda_{2}}{8} (\rho^{2}-f_a^2)^2 +\frac{1}{2}m^2_\beta \beta^2+\frac{1}{4} \lambda_{5} h^{2} \rho ^{2} \cos (2 \theta)+\mu \beta \rho \cos(\theta) \bigg).
\ee
In the following, the  case $\lambda_5=0$ will be considered, the effect of this parameter will be analyzed later on. Models of the type described above were considered recently in \cite{ketov}.

Before to enter in the details of the model it may be convenient to describe how the bound $f_a<10^{12}$GeV is avoided. Assume that $\rho$ rolls fast to its mean value $\rho=f_a$ inside the inflationary period while the  field $\chi$ drives inflation. The behavior of the field $\beta$ is not of importance, and may be slow rolling and subdominant. However, it should roll to its minima before the QCD era. The relevant point is the value of the parameter $\mu$, which should be small enough for the axion $a=f_a\theta$ to be frozen till the QCD era. In addition, the mass of the field $\beta$  should be $m_\beta>> H_{qcd}$, which insures that this field rolls from its initial value $\beta_0\sim M_p$ to its minima $\beta_m$ before the QCD era. The  minimum $\beta_m$ for a generic value of the axion $a$ can be calculated from (\ref{i}), the result is
$$
\beta_m=-\frac{\mu f_a}{m^2_\beta}\cos\bigg(\frac{a}{f_a}\bigg).
$$
In the last formula, it has been assumed that $\rho$ reached the minimum $\rho\sim f_a$. In these terms the part of the potential  (\ref{i}) corresponding to $\beta$ and $a=f_a\theta$ becomes
\be\lb{mn}
V(a)=-\frac{\mu^2 f_a^2}{2m^2_\beta} \cos^2\bigg(\frac{a}{f_a}\bigg).
\ee
On the other hand, if $\mu<<H^2_{qcd}$ the axion never moves, since its mass is smaller than the Hubble constant $H$ till the QCD era. Initially it was in a maximum $a\sim 0$. However, when $\beta$ went into a  minima, it follows
from (\ref{mn}) that the point $a\sim 0$ becomes a minima due to the appearance of the minus sign. But  since $a$ never rolled it is clear that its initial value at the QCD era is $a\sim 0$. This contradicts
the hypothesis of \cite{abbott} that $a\sim f_a \pi$ at the QCD era, thus the bound $f_a<10^{12}$GeV is neatly avoided. This is precisely the goal of the present work.

In addition to the features described above, it would be desirable to keep the standard  QCD axion description almost unchanged, and this impose further constraints for the parameter $\mu$. Recall that, near the QCD era, the standard temperature dependent axion mass $m_a(T)$ is turned on
and the axion potential in our model becomes
\begin{equation}\label{efecto3}
V(a)= -\frac{\mu^2 f_a^2}{2m^2_\beta} \cos^2\bigg(\frac{a}{f_a}\bigg)+m_a^2(T)f_a^2\bigg[1-\cos\bigg(\frac{a}{f_a}\bigg)\bigg].
\end{equation}
The axion mass $m_a(T)$ is the temperature dependent QCD one, its explicit form is \cite{axiontemp}
\be\lb{masaaxio}
m_a(T)\sim m_a(0)b\bigg(\frac{\Lambda_{qcd}}{T}\bigg)^4, \qquad T> \Lambda_{qcd},
\ee
with $b$ a model dependent constant. The mass $m_a(0)$ is the axion mass for temperatures $T< \Lambda_{qcd}$, it is temperature independent and its value is given by \cite{axion3}
\be\lb{aire}
m_a(0)\sim \frac{m_\pi f_\pi}{f_a}\sim 10^{-21}GeV.
\ee
The constraint to be imposed is  that the effect of the $\cos^2(a/f_a)$ to be smaller than the $\cos(a/f_a)$ one. In other words, the idea is not to modify standard QCD axion picture considerably. This will be the case when
$$
\mu^2<<m_a^2(0)m_\beta^2.
$$
Although the expect mass axion (\ref{aire}) is expected to be very tiny, the field $\beta$ is allowed to have mass values $m_\beta$ not far from the GUT scale, so $\mu$ may take intermediate values of the order of the eV$^2$ or MeV$^2$.

The last two paragraphs assume that $\chi$ is slow rolling and that $\rho$ rolls to its minima inside inflation. In order to further justify this assumption, assume that  both fields are slow rolling have initial transplanckian values $\rho_0$ and $h_0$ of the same order.  By taking into account the definition (\ref{aeb}) it is obtained that the contribution to $H^2$ of the field $h$, under the slow rolling assumption is
\be\lb{hh}
H_h^2=\frac{V_{hE}}{M^2_p}\sim \frac{\lambda_1 M_p^2}{\xi_1^2}\bigg(1-e^{-\sqrt{\frac{2}{3}}\frac{\chi}{M_p}}\bigg)^2.
\ee
On the other hand, as the value $f_a\sim M_p$ it follows that the $\rho$ contribution to the Hubble constant is
\be\lb{hr}
H_\rho^2=\frac{V_{\rho E}}{M^2_p}\sim (c^4-1) \frac{e^{-\sqrt{\frac{2}{3}}\frac{\chi}{M_p}}\lambda_2 M_p^2}{\xi_1^2}.
\ee
with the constant $c$ defined through $\rho=c f_a\sim c M_p$. This constant takes moderate values, of the order between the unity and $10^2$. Now, the kinetic plus the mass term for $\rho$ in (\ref{kined})  becomes
$$
L_{k\rho}= \frac{(\partial _{\mu} \rho)^{2}}{2\bigg(1+\frac{\xi_{1} h^{2}}{M_p^2}\bigg)}+\frac{\lambda_2 v_2^2\rho^2}{\bigg(1+\frac{\xi_{1} h^{2}}{M_p^2}\bigg)^2},
$$
where again (\ref{aeb}) has been taken into account. The kinetic term of the last expression is not canonically normalized. The canonical normalized field 
$$
\rho'=\frac{\rho}{\sqrt{1+\frac{\xi_{1} h^{2}}{M_p^2}}},
$$
acquires the following mass
\be\lb{mr}
m^2_{\rho'}=\frac{\lambda_2 v_2^2\rho^2}{1+\frac{\xi_{1} h^{2}}{M_p^2}}=e^{-\sqrt{\frac{2}{3}}\frac{\chi}{M_p}}\lambda_2 M_p^2.
\ee
This mass is to be compared with $H^2_h$ in (\ref{hh}) or $H^2_\rho$ in (\ref{hr}). When it is larger than the Hubble constant, the slow rolling condition for $\rho$ is spoiled. By comparing (\ref{hr}) and (\ref{mr}) it follows that, when $c^4<\xi_1^2$, one has that $m_\rho> H_\rho$.  In addition, when $\lambda_2 M_p^2>>\lambda_1 M_p^2\xi_1^{-2}$ it is seen by comparison (\ref{hh}) and (\ref{mr}) that, at the stages
$\chi\sim M_p$, the following inequality takes place
$$
m^2_{\rho'}>H^2_h.
$$
This shows that the assumption that $\rho'$ is slow rolling during inflation is not quite right. It is reasonable to assume that the Peccei-Quinn radial field $\rho$ in fact goes to its mean value $\rho=f_a\sim M_p$ during inflation while $\chi$ keeps the universe accelerating, as in ordinary Higgs inflation \cite{higgsinf}.

There exist scenarios with two fields evolving during inflation, for which one of the fields might roll quickly to the minimum of its potential and then the problem reduces to single field inflation. Models of hybrid inflation \cite{7} or other models of first-order inflation \cite{8}-\cite{9} provide examples of this situation. The analogous holds for the model presented here. Since the Peccei-Quinn symmetry is broken inside inflation the topological defects that may be formed are arguably diluted away by the rapid universe expansion. This point will be discussed in detail in the next section. Now, as the Peccei-Quinn rolls fast to the minima,  the dominant contribution for $H^2$ is $h$. Thus, the same cosmological bounds for $\xi_1$ as in standard Higgs inflation \cite{higgsinf} may be imposed as approximations namely, $\xi_1\sim 5 .10^4$ and $H_{inf}=\lambda_1M_p\xi^{-1/2}_1\sim M_{gut}$.

\subsection{Detectability of the $\beta$ scalar}

The previous scenario introduces a field $\beta$ which has a wide mass range  $H_{qcd}<m_\beta<M_{gut}$.  In view of this, it is of importance to discuss if this particle
can be detected in future colliders. This aspect may be clarified by analyzing its couplings to the other states of the model. An inspection of the potential (\ref{guha2}) shows that 
it has a coupling with the axion field $a$ and it mixes with the Peccei-Quinn field $\varphi$. This mixture is very small and will be analyzed below. 
As is well known, the hidden Higgs $\varphi$ in the KSVZ model is coupled to some hidden quark $Q$ which is singlet under the electroweak interaction \cite{ksvz}.
This coupling is given by 
\begin{equation}\label{ax}
 \mathcal L_{add}= i\overline{\psi} \gamma^\mu D_\mu \psi -(\delta \overline{\psi}_R \varphi \psi_L +\delta^\ast \overline{\psi}_L \varphi^\ast \psi_R)  \;.
\end{equation}
Here $\psi$ is the wave function of the hidden quark $Q$. The first term $i\overline{\psi}\gamma^\mu D_\mu \psi$ includes the kinetic energy of the new quark and its coupling with the gluons; the parameter $\delta$ of the Yukawa coupling between $\varphi$ and $\psi$ is an undetermined one. The heavy quark  mass is given by $m_\psi=\delta\varphi_0$. Note that the axion coupling constant is related to the vacuum expectation value according to $f_a=\sqrt{2}\varphi_0$, and the axion mass goes as $m_a\sim f_a^{-1}$. On the other hand, the mass of the quark $Q$ is proportional to $f_a$; so the heavier the quark is, the lighter the axion will be. The mass of the hidden quark is expected to be very large, since in our model $f_a\sim M_p$. A reasonable but not unique value may be that $m_Q\sim M_{gut}$, and we will use 
this value for estimations in the following. 
\begin{figure}[h!]
\centering
\includegraphics[width=0.5\textwidth]{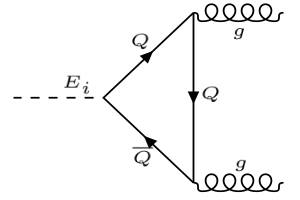}
$$
$$
\caption{Decay of the mass eigenstate $E_2$ into two gluons $G_\mu$.}
\label{figura1}
\end{figure}

Now, if the field $\beta$ is produced in an accelerator
then it may decay into the channel $\beta\to a+a$ or into two gluons by the triangle diagram of the figure \ref{figura1}.  Let us focus in this triangle diagram first.
The potential (\ref{guha2}) implies that $\beta$ and $\varphi$ mix, their mass matrix is
\begin{equation}
M=\left(
\begin{array}{cc}
  m^2_{\varphi} & \mu  \\
  \mu & m^2_\beta  
\end{array}
\right)
\end{equation}
The parameter $\mu$ is very small, namely $\mu<< m_\beta^2<<m_Q^2$, so the mass eigenvalues are essentially $m^2_\varphi$ and $m^2_\beta$.  The mass eigenstates are
then approximated by
$$
E_1\simeq \delta \varphi-\frac{\mu}{f_a^2}\delta \beta,\qquad 
E_2\simeq \delta \beta+\frac{\mu}{f_a^2}\delta \varphi.
$$
Here $\delta\varphi$ are the radial excitations of the field $\varphi$ and $\delta \beta$ the vacuum excitations of the $\beta$ field.
The first eigenstate corresponds to the mass $m_\varphi$ and the second one to $m_\beta$. The small mixing triggered by $\mu$ induces a Yukawa coupling for the  state $E_2$ with numerical value $\delta_{eff}\sim\mu\delta f_a^{-2}$. On the other hand, this second state is allowed to have a wide mass range, in particular, it may be  $m_\beta\sim 100$GeV, which is inside the current 
accelerator technology. The decay width of the diagram \ref{figura1} can be estimated in the limit $m_Q>>m_\beta$ as 
\be\lb{min}
\Gamma_2\simeq \frac{\delta_{eff}^2\alpha_s^2 m_\beta^3}{m_Q^2}\simeq \bigg(\frac{\mu}{f_a^2}\bigg)^2\frac{\delta^2\alpha_s^2 m_\beta^3}{m_Q^2}.
\ee
This value follows from dimensional analysis and from the fact that such decays are proportional to $m_\beta^3$ \cite{hun1}-\cite{hun4}.  If this were the main decay channel and we assume that the accelerator 
can reach the TeV scale, then the maximum probability of decay corresponds to $m_\beta\sim$TeV. The mean life time will be then
$$
\tau_2 \simeq \bigg(\frac{f_a^2}{\mu}\bigg)^2 \frac{m_Q^2}{\delta^2 \alpha_s^2 m_{\beta}^3} \geq 10^{30} yrs.
$$
Here it was assumed that $\alpha_s\sim 1$ and $\delta\sim 10^{-3}$.  This life time is enormous. The reason is that the triangle is very massive, and the coupling between $E_2$ and the fermions is of order $\mu/f_a$, which is extremely small. Thus, if the state $E_2$ were produced in an accelerator, its main decay channel would be $E_2 \to a + a$, which is faster than the triangle diagram channel. However, for this decay to take place, the state $E_2$ has to be produced inside the accelerator. A simple though shows that its main production channel  is given by
gluon fusion. This process is described by a diagram analogous to the one in figure 1. The cross section is given by \cite{hun1}-\cite{hun4}
$$
\sigma(gg\to \beta)=\frac{8\pi^2\Gamma_2}{N_g^2 m_\beta}\delta(s-m_\beta^2)
$$
where $\Gamma_2$ is given in (\ref{min}) and $N_g$ is the number of different gluons. It follows then from (\ref{min}) that
$$
\sigma(gg\to\beta)\sim\bigg(\frac{8\pi^2\mu}{f_a^2}\bigg)^2\frac{\delta^2\alpha_s^2 m_\beta^2}{N_gm_Q^2}.
$$
This expression is fully suppressed since $m_\beta<<m_Q$ and $\mu<<f_a^2$. Thus, the state $E_2$ can not be produced in a modern accelerator and is not dangerous from the phenomenological point of view.

\section{The issue of topological defects formation}

In the previous sections, a model that solves the isocurvature between axion and high energy inflationary models  has been constructed. In addition, it has been shown
that, for this scenario,  the vacuum realignment mechanism does not give a significant contribution to the present energy density. However, there exist other possible sources of axions namely, topological defects.
In fact, this issue is a delicate one, since a  density value large enough of such defects may be 
in direct conflict with observations.  In the following this problem will be considered in certain detail. The analysis to be done below  is based in some standard references such as \cite{harari}-\cite{harari3} and \cite{topo1}-\cite{topo10}, where some numerical features are largely discussed.

\subsection{Generalities about defect formation}

 It may be convenient  to discuss first some general knowledge about topological defects formation,
this knowledge will be applied to our specific case later on.
\\

\emph{Axion production by global strings:} Consider first the simplest Peccei-Quinn model 
$$
L=\frac{1}{2}\partial_\mu \Phi \partial^\mu \Phi^\ast+\frac{\lambda}{2}(\Phi\Phi^\ast-f_a)^2.
$$ 
The global U(1) transformation $\Phi\to e^{i\alpha}\Phi$  is  a symmetry for the model. This scenario admits cosmic strings for which the mean value $<\Phi>$ is different from $f_a$ only inside the string core.
The width of the core is of the order $\delta_s\sim (\lambda f_a)^{-1}$. For a long $n=1$ strings one has $<\Phi>\sim f_a e^{i\phi}$ outside the string core, with $\phi$ the azimutal angle and the string is assumed to lie on the $z$ axis.
The energy of such strings is divergent, since
the $U(1)$ symmetry of the model is a global one. However, a natural cutoff is the typical curvature radius of the string or a typical distance between two adjacent
strings. By denoting such cutoff as $L$ it follows that the energy per length of the string is
$$
\mu\sim f_a^2\log(L f^{-1}_a\lambda^{-1}).
$$
Two strings with different values of $\Delta \theta$ attracts one to another with a force $F\sim \mu/L$. The scale of the string system at cosmic time $t$ is of the order of $t$.

The number of strings inside every horizon is of course an unknown parameter. However, it is plausible that the values of the axion $a(x,t)$ are uncorrelated
at distances larger than the horizon. If this is the case, then by traveling around a path going through a path with dimensions larger than the horizon  size
one has that $\Delta a=2\pi f_a$. This suggest the presence of an string inside any horizon zone. These strings are stuck into a primordial plasma and their density grow
due to the universe expansion $a(t)\sim \sqrt{t}$. However, the expansion dilutes the plasma and at some point, the string starts to move freely.
The energy density of strings is know to be $\rho_s\sim\mu/t^2$. 
For matter instead, such density is $\rho_m\sim 1/G_N t^2$ \cite{topo3}. The quotient between these contributions is
$$
\frac{\rho_s}{\rho_m}\sim \bigg(\frac{f_a}{M_p}\bigg)^2\log\bigg(\frac{t}{\lambda f_a}\bigg).
$$
The density of axions produced by these strings has been calculated in \cite{topo1}, the result is roughly
\be\lb{disa1}
n_a^{s}(t)=\frac{\xi r N^2}{\chi}\frac{f_a}{t^2},
\ee
where $\xi$ is a parameter of order of the unity, and the other unknown parameters $\chi$ and $r$ take moderate values. In particular, the parameter $\chi$ express our ignorance about the precise value of the cutoff $L$.
The contribution to the energy density coming from these strings is 
\be\lb{disa6}
\rho_s=m_a\frac{L r}{\chi}\frac{N^2 f_a^2}{t_1}\bigg(\frac{a_1}{a_0}\bigg)^3.
\ee
Here $a_1/a_0$ is the quotient between scale factor at the time $t_1$ and the present one. This density should not be larger that the critical density today, and this requirement usually impose constraints for the models on consideration.
\\

\emph{Defects produced by massive axions:} The other case to be considered is that the Peccei-Quinn symmetry is only approximated, which means that the axion is massive from the very beginning \cite{topo3}. In several axion models, this picture holds for times larger than the age $t_1$ defined by  $m_a(t_1)t_1=1$. However, in our case the axion is massive from the very beginning. Now, in a generic situation,
by assuming that the radial oscillations of the Peccei-Quinn filed are not  large enough, the effective lagrangian for the $\theta$ field is
$$
L_s=f_a^2\partial_\mu \theta \partial^\mu \theta+m_a^2(\cos\theta-1).
$$
The equation of motion derived from this lagrangian is
$$
\partial_\mu \partial^\mu \theta+m_a^2\sin\theta=0.
$$
A domain wall solution for this equation is
\be\lb{disa2}
\theta=4\tanh^{-1}\exp(m_a x),
\ee
where $x$ is the direction perpendicular to the wall. The thickness of the wall is approximately 
$\delta\sim \frac{1}{m_a}$. The energy density per unit area is exactly 
$\sigma=16 m_a^2 f_a$. These defects are formed as $m_a>t^{-1} $. At later time the system corresponds to strings connected by domain walls.
Their linear mass density is
\be\lb{disa3}
\mu\sim f_a^2\log(m_a \lambda f_a)^{-1}.
\ee
These strings form the boundary of the walls and of the holes in the wall.  
The particles and strings does not have an appreciable friction on the wall. The force tension for an string of curvature $R$ is
$F\sim\frac{\mu}{R}$, and this quantity is smaller  than the wall tension $\sigma$ when
\be\lb{disa4}
R<\frac{\mu}{\sigma}.
\ee
At  $t<\mu/\sigma$ the evolution is analogous to the massless case. In the opposite case $t>\mu/\sigma$, the physics goes as follows. The curvature radius $R$ becomes large and the system is dominated by the wall tension.
The domain walls will shrink and pull the strings together. As the wall shrinks, their energy is transferred to the strings, and energetic strings pass one into another and the walls connecting them shrinks. As a result the system violently oscillates and intercommute. Due to this behavior, the strip of domain wall connecting the intercommuting string breaks into pieces. When the intersection probability is $p\sim 1$ the strings break into pieces
$\mu/\sigma$ at $t\sim \mu/\sigma$. A piece of wall of size $R$ losses its energy due to oscillations as
$$
\frac{dM}{dt}\sim -GM^2 R^4 \omega^6-G\sigma M.
$$
The decay time is
\be\lb{disa5}
\tau\sim \frac{1}{G\sigma}.
\ee
For closed strings and infinite domain walls without strings the mean life time is of the same order. This result is independent on the size, thus the domain walls disappear shortly.
The contribution to the energy density is 
\be\lb{disa7}
n_a^{s}(t)=\frac{6 f_a^2}{\gamma t_1}\bigg(\frac{R_1}{R_0}\bigg)^3.
\ee
Here $\gamma$ is an unknown parameter which in numerical simulations seems to be close to $7$.  Usually the domain walls contributions (\ref{disa7}) are subdominant with respect to the string contributions.

\subsection{The formation of defects in our model}

After discussing these generalities, the next point is to analyze the presence of topological defects in the our model. At first sight, the axion we are presenting is massive at the early universe and the direct application of  (\ref{disa7})
with $f_a\sim M_p$ gives an unacceptably large value for the energy density. However, as discussed below (\ref{disa2}), the domain walls are formed when $m_a>t^{-1} $. This arguably never happens in our case since the axion mass is  fixed to be $m_a<H\sim t^{-1}$ until the very late time $t_1 m_a(t_1)=1$.  On the other hand by defining the "string" time
$$
t_c=\frac{\mu}{\sigma}\simeq\frac{f_a}{m_a^2},
$$
it follows that the condition $t>\mu/\sigma$ is not satisfied until  the universe age  is close to $t_1$.  Before this era, as argued below (\ref{disa4}) the massless string description is the correct one. The direct application of  the formula (\ref{disa6}), which is valid for the massless case, also gives a bad result. However, in our case, the symmetry breaking occurs inside the inflationary period. Thus the argument that there is at least one string per horizon given above  formula (\ref{disa1}) is not necessarily true, instead the axion value is arguably homogenized over an exponentially large region, and the strings are diluted away. The standard picture is that when $t=t_1$ the strings are edges of $N$ domain walls, but we expect this dilution to be such that the radiated axion density is not significant. Of course, a precise numerical simulation for this may be very valuable in a future. In any case, our suggestion is that the defects that appear in our scenario are not dangerous from the cosmological point of view due to the mentioned dilution.

\section{The consequences of the parameter $\lambda_5$ of the model}
In the previous sections, the parameter $\lambda_5$ has been set to zero in (\ref{i}). One of the reasons is that a non zero value for this parameter induces a term proportional to $\cos(2\theta)$ for the axion.
The factor 2 inside this cosine is problematic. Recall that our model, as customary in axion physics, assumes that the axion $a=f_a\theta$ is initially at a maximum. But, due to the factor 2, this maximum may be $a\sim 0$ as before, or $a\sim \pi$. If the parameter
$\lambda_5$ is small enough, then the axion is frozen till the QCD era. Near this era the term $V_{qcd}(a)= m^2_a(T)f_a^2(1-\cos\theta)$ is turned on. The value $a\sim 0$ becomes a minima when this term appears. However, it is simple to check that the value $a\sim \pi$ is still a maximum. The last situation is inside the hypothesis of \cite{abbott}, thus the misalignment mechanism produces an extremely large value for the axion density today. This density is larger than the critical density, and this do not pass cosmological tests.

In addition, note that the $\lambda_5$ part of the potential (\ref{i}) at the reheating period, for which $\chi\sim 0$, is 
$$
V(\theta, h)=\frac{1}{4} \lambda_{5} h^{2} \rho^{2} \cos (2 \theta),
$$
where the overall exponential $e^{-2\sqrt{\frac{2}{3}} \frac{\chi}{M_p}}$ has been neglected since $\Omega^2\sim 1$.
By taking into account (\ref{aeb}) it follows that \footnote{Note that the quantity $\chi$ is  replaced by $|\chi|$. This distinction is not essential during inflation but it is during the reheating period 
\cite{higgsinf}.}
$$
h^2\sim \frac{M_p^2}{\xi_1}(1-e^{-2\sqrt{\frac{1}{6}} \frac{\chi}{M_p}})e^{-2\sqrt{\frac{1}{6}} \frac{\chi}{M_p}}\sim \frac{M_p|\chi|}{\xi_1},
$$
for very small $\chi$.
Thus there is a coupling between the axion $a$ and the Higgs related field $\chi$ of the form
$$
V(\theta, h)=\frac{\lambda_5M_p|\chi|}{4\xi_1}f_a^2 \cos\bigg(\frac{2a}{f_a}\bigg),
$$
which generates at first order a Yukawa coupling mass term
\be\lb{ma}
L_Y=\frac{8\lambda_5M_p|\chi|}{\sqrt{6}\xi_1} e^2, \qquad m^2_a=\frac{8\lambda_5M_p|\chi|}{\sqrt{6}\xi_1}
\ee
with $e=\pi-a$ the axion fluctuation from its initial minima $a=\pi$. The oscillations of $\chi$ may induce non perturbative generation of axions \cite{higgsinf}. In fact, the equation of motion for the $k$ Fourier component $e_k$ is then
$$
\frac{d^2e_k}{dt^2}+\bigg(\frac{k^2}{a^2}+m_a^2\bigg)e_k=0.
$$
This equation of motion is formally identical to the one for the vector bosons $W_k$ considered in \cite{higgsinf}. This reference shows that during the reheating period the scale factor goes as $a(t)\sim t^{2/3}$
for $t$ the coordinate time, and corresponds to a matter dominated period. In addition, the time behavior of the field $\chi$ is approximated by 
$$
\chi(t)\sim \frac{\chi_{end}}{\pi j}\sin(M t),\qquad M=\frac{M_p}{\xi_1},\qquad \chi_{end}\sim M_p.
$$
The non perturbative creation of particles takes place in the non adiabatic period for which $m_a$ satisfy
$$
|\frac{dm_a}{dt}|>m_a^2.
$$
In this region one may use the approximation $\sin(M t)\sim M t $ and the equation of motion becomes
$$
\frac{d^2e_k}{d\tau^2}+\bigg(K^2+|\tau|\bigg)e_k=0,
$$
where the following quantities 
$$
\tau=\gamma t,\qquad \gamma=\bigg(2\frac{M_p\lambda_5\chi_{end}M}{\sqrt{6}\pi j \xi_1}\bigg)^{1/3},\qquad K=\frac{k}{a\gamma},
$$
have been introduced, with $a(t)$  taken as a constant for each oscillation. Since this equation is already considered in \cite{higgsinf} we can take as granted the results of that reference. In these terms, it is found that the number of axions generated in the first oscillations is
\begin{equation}
n(j)=\frac{1}{2 \pi ^{2} R^{3}} \int_{0}^{+\infty} dk k^{2} [|T_{k}|^{2}-1]=\frac{q_{a}}{2} I M^{3},
\label{pp25}
\end{equation}
with $$I=0.0046,\qquad q_a=\frac{M_p \lambda_5 \chi_{end}}{ \xi_1 \pi M^2}.$$ Since $\xi_1 \sim 5 \cdot 10^4$ it follows that, in the first oscillation, the mean axion number is $n_a^{(1)} \sim \lambda \cdot 10^{46}$GeV$^3$. The averaged mass during the first oscillation is given by $$m_a^{(1)} \sim \frac{2 M_p \lambda_5 \chi_{end}}{ \xi_1 \pi} \sim \lambda_5 \cdot 10^{34}GeV.$$ Thus the axion density present at this early stage is
 $$\rho_a^{(1)}\sim m_a^{(1)} n_a^{(1)} \sim \lambda_5^2 10^{80} GeV^4.$$ From here it follows that for a value $\lambda_5 \sim 10^{-7}$ the density value is around $\rho_a^{(1)} \sim 10^{66}$ GeV$^4$ which is two orders less than the critical density at this stage namely, $\rho_{c} \sim 10^{68} $GeV$^{4}$. 

 The value $\lambda_5<10^{-7}$ is small but reasonable. However, the addition of the $\lambda_5$ term $V_5=\lambda_5\Phi^2 \varphi^2$ may generate a large mass term for the Higgs $\Phi$ when the Peccei-Quinn field $\varphi$ goes to its mean value $\Phi \sim f_a e^{i\theta}\sim M_p e^{i\theta}$. The resulting additional mass is of the form $m'_h\sim \lambda_5 f_a^2\sim \lambda_5 M_p^2$. This term should not affect the ordinary Higgs mass term and this condition forces $\lambda_5<10^{-34}$. This value is extremely small and the resulting energy density is suppressed at least by $27$ orders of magnitude from the critical one.

For all the reasons stated above, it is safe to assume that the  $\lambda_5$ term should be strongly suppressed and, in fact, $\lambda_5$ may be set equal to zero.

\section{Discussion}

The present work introduces a two field  inflationary model involving the KSVZ Peccei-Quinn hidden Higgs $\varphi$ and the ordinary Higgs $\Phi$. This model is in agreement with the basic cosmological constraints, and relaxes the tension between axion and high energy inflationary models. Furthermore, the presence of small but explicit Peccei-Quinn violating terms induce a non zero axion potential $V(a)$, whose sign is opposite to the standard one $V(a)\sim m_\pi^2 f_\pi^2(1-\cos(a/f_a))$. This interchanges minima with maxima at some point of the universe evolution and in particular, the point $a=0$ is a maxima at the beginning of the universe. This suggest that if here that the dynamic of such axion is such that $m_a(t)<H(t)$ then it never rolls from the top of the potential $a\sim 0$. At large times the contribution $m^2_\pi f_\pi^2$ is turned on, and the potential changes the sign. However the axion did not roll and it stays now near the minima $a=0$. Under these circumstances the bounds of \cite{abbott} are avoided. Thus the axion constant is not forced to be $f_a<10^{12}$GeV. In fact  it can be $f_a\sim M_p$, which is in harmony with high scale inflationary models. This implies that the axion mass may be  $m_a\sim m_\pi f_\pi/f_a\sim 10^{-21}$GeV, which is a very tiny value.

The model presented here introduces a real scalar $\beta$ which may have a mass below the TeV scale. However, we have checked that this state is sterile from the accelerator point of view, since its coupling with the Standard Model particles is strongly suppressed.

We have also discussed the formation of topological defects for the present model. Although our discussion is not numerically precise, we suggest that the density of these defects is not considerable and they do not constitute a problem from the cosmological point of view. 

In the authors opinion, the model presented here complements the ones of the reference \cite{doublen}, which corresponds to the DFSZ axion model. In the later case, the isocurvature problem is  avoided in the context of several Higgs inflationary scenarios \cite{doublen}. The model that the authors of \cite{doublen} consider contains
three Higgs $H_u$, $H_d$ and $\phi$. The two fields $H_u$ and $H_d$ are coupled to the curvature with couplings $\xi_u$ and $\xi_d$, while $\phi$ is not.
All these three fields have Peccei-Quinn charges, and the axion $a$ is identified as a combination of the phases $\theta_u$, $\theta_d$ and $\theta_\phi$ for these fields. The coefficients of this combination depend
on the mean values of these fields. The main point  is that the mean effective values of the radial fields have different values at the inflationary epoch than today. At present universe these mean values satisfy $v_\phi>>v_u, v_d$  and the axion today is given predominantly by the phase $\phi_\theta$.  At the early universe instead,  $v_\phi<<v_u, v_d$ and the axion mainly a combination of the phases $\theta_u$ and $\theta_d$.  Based on this the authors of \cite{doublen} construct an ingenious mechanism for which the isocurvature fluctuations are strongly suppressed. 

The model described above for the DFSZ axion relies in a mixing of  phases, while there is only one phase in the standard KSVZ axion model. Thus, the techniques employed in \cite{doublen} are not applied directly to this axion model.
This is in part one of the motivation for searching for alternative mechanisms such as the ones presented here.

There is an aspect our model that deserves, in our opinion, to be improved. The fact that the axion is performing small oscillations on the minima of the potential $V(a)$ implies that it contribution to the present energy density of the universe is not appreciable. Thus, our model solves the isocurvature tension between the high energy inflationary models and the axion ones, but at cost of discarding the axion as the main component of dark matter.
It may be interesting to find variation of our scenario where the predicted density is of the order of the critical one, but we suspect that this is not an easy task.
We leave this for a future research.  
\\

{\bf Acknowledgements:}  Some discussions with R. Sassot, D. de Florian, E. Alvarez and D. Lopez-Fogliani are acknowledged. The authors are supported by CONICET (Argentina).

\end{document}